\documentclass[twocolumn]{aastex631}

\begin{document}

\title{HWO Target Stars and Systems: A Prioritized Community List of Potential Stellar Targets for the Habitable Worlds Observatory's ExoEarth Survey}

\author[0000-0003-3989-5545]{Noah W.\ Tuchow}
\affiliation{Steward Observatory and Department of Astronomy, The University of Arizona, Tucson, AZ 85721, USA}
\affiliation{NASA Goddard Space Flight Center, Greenbelt, Maryland, USA}
\email{nwtuchow@arizona.edu}

\author[0000-0001-5737-1687]{Caleb K.\ Harada}
\affiliation{Department of Astronomy, 501 Campbell Hall \#3411, University of California, Berkeley, CA 94720, USA}

\author[0000-0003-2008-1488]{Eric E.\ Mamajek}
\affiliation{Jet Propulsion Laboratory, California Institute of Technology, 4800 Oak Grove Dr., Pasadena, CA 91109, USA}

\author[0000-0002-2903-2140]{Angelle Tanner}
\affiliation{Mississippi State University, 355 Lee Hall, Starkville, MS, USA}

\author[0000-0003-0595-5132, gname=Natalie, sname=Hinkel]{Natalie R.\ Hinkel}
\affiliation{Louisiana State University, Department of Physics and Astronomy, 202 Nicholson Hall, Baton Rouge, LA 70803, USA}

\author[0000-0002-4951-8025]{Ruslan Belikov}
\affiliation{NASA Ames Research Center, Moffett field, CA 94035, USA}

\author[0000-0001-8795-0110]{Dan Sirbu}
\affiliation{NASA Ames Research Center, Moffett field, CA 94035, USA}

\author[0000-0002-5741-3047]{David R.\ Ciardi}
\affiliation{NASA Exoplanet Science Institute-Caltech/IPAC, Pasadena, CA 91125, USA}
 
\author{Christopher C.\ Stark}
\affiliation{NASA Goddard Space Flight Center, Greenbelt, Maryland, USA}

\author[0000-0002-4852-6330]{Rhonda M.\ Morgan}
\affiliation{Jet Propulsion Laboratory, California Institute of Technology, 4800 Oak Grove Dr., Pasadena, CA 91109, USA}

\author[0000-0002-8711-7206]{Dmitry Savransky}
\affiliation{Sibley School of Mechanical and Aerospace Engineering, Cornell University, Ithaca, NY 14853}

\author[0000-0002-6463-063X]{Michael Turmon}
\affiliation{Jet Propulsion Laboratory, California Institute of Technology, 4800 Oak Grove Dr., Pasadena, CA 91109, USA}

\begin{abstract}
The HWO Target Stars and Systems 2025 (TSS25) list is a community-developed catalog of potential stellar targets for the Habitable Worlds Observatory (HWO) in its survey to directly image Earth-sized planets in the habitable zone. The TSS25 list categorizes potential HWO targets into priority tiers based on their likelihood to be surveyed and the necessity of obtaining observations of their stellar properties prior to the launch of the mission. This target list builds upon previous efforts to identify direct imaging targets and incorporates the results of multiple yield calculations assessing the science return of current design concepts for HWO.  The TSS25 list identifies a sample of target stars that have a high probability to be observed by HWO (Tiers 1 and 2), independent of assumptions about the mission's final architecture. These stars should be the focus of community precursor science efforts in order to mitigate risks and maximize the science output of HWO. This target list is publicly available and is a living catalog that will be continually updated leading up to the mission.
\end{abstract}

\section{Introduction}

NASA's upcoming Habitable Worlds Observatory (HWO) has the ambitious goal of directly imaging Earth-sized planets in the habitable zones of Sun-like stars (exoEarths).  
It aims to meet the science goals posed by the Astro 2020 Decadal Survey to detect 25 exoEarths and spectrally characterize their atmospheric compositions \citep{DecadalSurvey, Feinberg24}.
To achieve these goals, HWO will use an onboard coronagraph to subtract out the light of a star and image faint planets in reflected light. 
From direct observations of planetary spectra, we hope to identify potential signs of life and better understand the processes that influence planetary habitability \citep[e.g.][]{Kopparapu20,Young24,Krissansen-Totton25}  

Because of the difficulty in obtaining the deep planet-star contrasts at close angular separations necessary to image planets in the habitable zone, only a select number of bright, nearby stars will be suitable targets for HWO's exoEarth survey (exoEarths are about $10^{10}$ times fainter than their host stars with angular separations on the order of 10-100 mas). 
Understanding this sample of neighboring stars is critical to ensure that the mission will be able to detect planets around them and to enable us to place meaningful constraints on their bulk properties, which in many cases are measured relative to the host star.
Therefore, these stars are prime targets for HWO precursor science studies - research required before the mission's design has been finalized, which informs the mission architecture and what science cases will be feasible.
Because of the key role these stars play in the success of the mission, the HWO Target Stars and Systems (TSS) sub-working group was established\footnote{The TSS sub-working group was itself part the of the broader Living Worlds Working Group organized by the HWO Science, Technology, Architecture Review Team (START).}.
The TSS group was tasked with compiling a list of stars that would be feasible targets for HWO, assessing the state of our knowledge about their stellar properties, and determining which observations are most essential to meet HWO's science goals and when such observations would be required.

\subsection{Motivation of the Catalogs and Databases Task Group}
\label{CDTG_motivation}
To deal with the first of these tasks - identifying a list of stellar targets for HWO's exoEarth survey - the TSS sub-working group organized the Catalogs and Databases Task Group (CDTG). In this study, we describe the work of the CDTG to construct the HWO Target Stars and Systems 2025 (TSS25) list, a prioritized catalog of probable stellar targets for HWO.
The goal in the construction of this target list was to generate a comprehensive list of HWO target stars, incorporating and building upon the results of previous catalogs and exoplanet yield studies. This TSS25 list was then provided to the other task groups organized by the TSS group (Fundamental Stellar Properties, High Energy Emission, Stellar Multiplicity, Activity and Rotation, and Radial Velocities), who were responsible for assessing the current availability of information in their areas of focus. These groups worked to determine what precursor observations and analyses are needed to inform HWO's design in the near-term, and what preparatory observations and analyses are required before launch to inform target selection and the analysis of HWO exoplanet spectra. 

The CDTG's work to identify the likely targets for HWO was made more difficult by the fact that HWO is still early in its development and does not yet have a finalized architecture. 
The design of the telescope and coronagraph can play a key role in determining what stellar targets are accessible to an exoEarth survey, with different designs able to target different samples of host stars \citep{Stark14}.  Conversely, the availability and characteristics of actual targets can shape the architectural needs of the telescope in order to achieve its science goals. 
{While HWO has not yet settled on a definitive architecture, its design is not totally unconstrained. There are a range of potential designs being explored in the form of the Exploratory Analytic Cases (EACs). These EACs are being studied in terms of their feasibility, technology maturity, expected cost, and their ability to meet HWO's science goals. These proposed architectures consider a range of telescopes diameters from 6 to 8 meters and a wide variety of coronagraph designs including parallel multi-channel coronagraphs \citep{Feinberg24}. These designs explore on-axis vs. off-axis coronagraph configurations, and investigate a broad variety of detector options ranging from mature \mbox{EMCCDs} to lower technology readiness level energy resolving detectors.}

Multiple catalogs of potential direct imaging targets for an exoEarth survey have been compiled in past studies, differing in purpose and assumptions about the mission design. 
For instance, \citet{Tuchow24} constructed the HWO Preliminary Input Catalog (HPIC), a catalog of $\sim$13,000 bright, nearby target stars that are potential targets for HWO. This input catalog is largely independent of assumptions about HWO's mission design, only making cuts based on distance and magnitude. It was developed to be used as an input for trade studies and yield calculations, so the sample of stars in the HPIC is sufficiently broad to allow for simulation of very optimistic telescope designs (i.e. 15m diameter LUVOIR A).
Taking a very different approach, the NASA Exoplanet Exploration Program's Mission Star List for HWO (hereafter the ``ExEP list")  ``{compile[d] a list of $\sim$160 stars whose exo-Earths would be the most accessible for a systematic imaging survey of habitable zones with a 6-m-class space telescope in terms of angular separation, planet brightness in reflected light, and planet-star brightness ratio}" \citep{Mamajek24}. 
The purpose of the ExEP list was to produce a {\it provisional} list of 100+ ``best" target stars to inform early community precursor science studies to improve the fidelity of yield simulations, understand the availability and limitations of observational data, and motivate community plans to propose for new observations on existing observatories, or propose new analyses of archival data. 

Both the ExEP and HPIC catalogs are useful for different purposes and complement each other. The HPIC makes minimal assumptions about the the design of HWO, causing it to be very large and contain many objects that wouldn't be good targets for a specified HWO design. 
The ExEP list on the other hand is much smaller, aiming for the subset of ``best" targets for a 6m diameter HWO, with the purpose to inform early precursor science efforts to improve our knowledge of prime HWO targets, but it makes assumptions about design of HWO that may differ from the final design. 

Beyond the existing target lists, the results of yield calculations provide additional constraints on the sample of stars that are plausible targets for direct imaging with HWO. 
Currently trade studies and yield calculations are ongoing to determine whether proposed mission designs will be able to meet the mission's science goal of characterizing 25 exoEarths \citep{DecadalSurvey}. 
Yield calculations use an input catalog of a large sample of stars and calculate the observability of planets around them with a given mission design \citep{Brown05,Savransky10}. Balancing the probability that a star hosts a detectable exoEarth and the exposure time required for spectral characterization, yield studies output a prioritized list of stars to observe to maximize the number of exoEarths detected \citep{Stark14}. The stellar population that yield calculations select to be surveyed is strongly dependent on one's choice of telescope and coronagraph design. For a given mission design, the results of yield calculations can inform which stars are the best targets based on required exposure time and the probability of detecting exoEarths.

Using the results of yield calculations alongside the ExEP list and the HPIC, the CDTG worked to synthesize all of this information into one comprehensive target list. We detail the construction of this target list in Section \ref{methods} and describe categorization of stars into different priority tiers. In Section \ref{t2_section} we describe how we identify a population of stars that are likely to be observed by HWO, using the results of several yield calculations encompassing the different HWO designs under consideration. 
We describe the specific yield calculations used in this study in Section \ref{yield_calcs}.
In Section \ref{data_products} we describe the contents of the TSS25 list.

\section{Construction of Priority Tiers}

\label{methods}
\begin{table*}[htb!]
    \centering
    \caption{Description of Tiers of HWO Targets}
    \begin{tabular}{l|c|l}
        \hline
        Tier & Number of Stars & Description \\
        \hline 
        1 & 164 & Most accessible targets for exoEarth direct imaging \\
        2 & 495 & Targets that could plausibly be observed by proposed HWO designs  \\
        3 & 12285 & Additional nearby, bright objects that are potential HWO targets \\
         
    \end{tabular}
    \label{tiers_table}
\end{table*}

The primary data product of the CDTG was the TSS25 list of target stars for HWO. 
In this catalog we define priority tiers of stellar targets according to their likelihood of being observed by HWO, their expected contribution to mission science yield, and the stellar properties required for each population. A summary of our priority tiers is shown in Table \ref{tiers_table}. The tiers are defined as follows: 

\begin{itemize} 
\item {\it Tier 1} is composed of the most promising targets for exoEarth direct imaging, requiring the lowest exposure times and with the highest probability of hosting detectable exoEarths.
\item {\it Tier 2} { consists of additional} stars that are plausible targets to be observed by HWO, taking into account the wide variety of architectures under consideration.
\item {\it Tier 3} {contains all remaining} nearby, bright stars that are potential targets for exoEarth direct imaging. 
\end{itemize}

One of the primary motivations for the separation of this target list into tiers is the fact that the different populations of stars that are potential HWO targets will require different precursor observations.
Different measurements of stellar properties will be needed for stars in each tier, and these will vary in terms of the degree of vetting and measurement precision required.
{For yield calculations, one requires a large input catalog of potential targets, but most of these are field stars which are unlikely to be selected for survey.
Coarse measurements of the properties of these stars (typically on the order 10-20\% uncertainty  depending on the specific property) are sufficient to give accurate yield results, so long there are not systematic biases at the population level \citep{Stark19,Tuchow24}.}
On the other hand, for the highest priority HWO targets, we want to obtain the most accurate and precise measurements of stellar properties and acquire difficult to obtain measurements that aren't available for the average star in a larger input catalog.
{These stars will be targeted in a search for planets, and since for most detection techniques the properties of the planet are measured relative to the star, we want to minimize the uncertainty in stellar properties to maximize the precision to which we can characterize their planets.}
The TSS group seeks to ensure that the highest priority objects don't host potential obstacles to direct imaging such as circumstellar disks and unseen binary companions.  Identifying which objects belong to each tier will allow the other task groups in the TSS working group to assess which measurements are necessary for which population of objects and determine what precursor observations are still required. {\it The TSS25 list is intended to be a living document}, and as we obtain additional observations of these target stars prior to the finalization of the mission, their priority tiers will be updated in future iterations.

\subsection{Tier 1}

Tier 1 of the TSS25 list consists of stars which have the most accessible habitable zones for direct imaging observations of exoEarths. 
It is composed of objects {that have} the highest probability of hosting detectable exoEarths and the shortest exposure times to characterize them. 
The exposure time required to characterize exoEarth atmospheres will range from a few minutes for a small number of ideal targets, while others will require upwards of a week.
{The exposure time to achieve a given signal-to-noise is influenced by several key factors including the brightness of the host star, the planet-to-star contrast, and the amount of exozodiacal dust, alongside the angular separation, ensuring that the planet is between the coronagraph's inner and outer working angles and that the coronagraph has sufficient throughput. }
Those stars requiring the shortest exposure times will be optimal targets for a direct imaging mission and will be frequently selected regardless of the mission design {(as shown by the yield calculations in Section \ref{yield_calcs})}. Tier 1 stars will likely result in the majority of the mission’s yield of Earth-like planets{. T}he remainder of the survey {would be} devoted to lower priority (Tier 2) targets{ which require} longer exposure times {and have} lower expected yields of exoEarths per star \citep{Tuchow24}.

To identify which objects belonged in Tier 1, we consulted the ExEP Mission Star List for HWO \citep{Mamajek24}. The ExEP list was constructed based on the observability of Earth sized planets in the habitable zone using a 6m diameter telescope. A sample of 164 high priority target stars were selected by \citet{Mamajek24} based on the angular separation and planet-star contrast of Earth-sized planets receiving the same stellar flux as the present day Earth. 
They assigned each of these objects to three tiers (A, B, and C) based on the angular separation and contrast requirements to detect exoEarths and the presence of potential obstacles to direct imaging such as stellar multiplicity and the presence of circumstellar material.  
Though the ExEP list makes some assumption about HWO's design (namely that it will have a 6m inscribed diameter), from the results of the yield calculations used to construct Tier 2 {in Section \ref{yield_calcs}}, we can be confident that the stars in the ExEP list are generally the most likely targets to be selected by any given HWO design. These stars are frequently selected due to their proximity to Earth and their brightness, resulting in exoEarths having a wider angular separation from their host star and a shorter required exposure time. 

In precursor science efforts for HWO, these high priority targets have received significant attention. 
For example, the SPORES-HWO (System Properties and Observational Reconnaissance for Exoplanet Studies with HWO) group has published a catalog of stellar properties of the 164 ExEP stars \citep{Harada24a}. The SPORES-HWO catalog includes fundamental stellar properties derived from SED fitting, stellar abundances, stellar variability, and high-energy emission information. The SPORES-HWO group has also investigated the state of precise radial velocity data of the ExEP stars, determining updated limits on planetary masses in those systems \citep{Harada24b}. SPORES-HWO has ongoing and future efforts including further radial velocity characterization of potential HWO target stars and ground-based imaging to place limits on close stellar companions.

\subsection{Tier 2 }
\label{t2_section}

Our goal in the construction of Tier 2 was to identify stars that have a high probability of being observed by HWO's exoEarth survey while making minimal assumptions regarding the mission architecture. {It consists of plausible HWO targets that aren't included in Tier 1.}  As the architecture for HWO has not been finalized, and the choice of telescope and coronagraph design has a strong influence on the observable stellar population \citep{Stark14}, identifying which stars belong in Tier 2 is challenging. 
{However, as described in Section \ref{CDTG_motivation}, the HWO project office is exploring multiple designs in the form of the EACs, which range in telescope diameter from 6 - 8m and investigate several possible coronagraph and detector options. }
Focusing on designs within {the trade space covered by the EACs} helps us place constraints on HWO's range of targets.

\subsubsection{Yield Calculations Used For Tier 2 Construction}
\label{yield_calcs}

To construct Tier 2, the CDTG was provided with results of the yield calculations of \citet{Stark24} and \citet{Morgan24}, covering mission designs spanning the expected trade space of HWO architectures.
Both of these results use different yield codes, AYO \citep[Altruistic Yield Optimization;][]{Stark14} and EXOSIMS \citep[Exoplanet Open-Source Imaging Mission Simulator;][]{Savransky2016,Savransky2017}, respectively. These yield frameworks implement different target prioritization schemes and often select different stars for survey. By using results from multiple yield codes, we will ensure that our results are not dependent on the assumptions of an individual yield calculation. 

\begin{figure*}
    \centering
    \plottwo{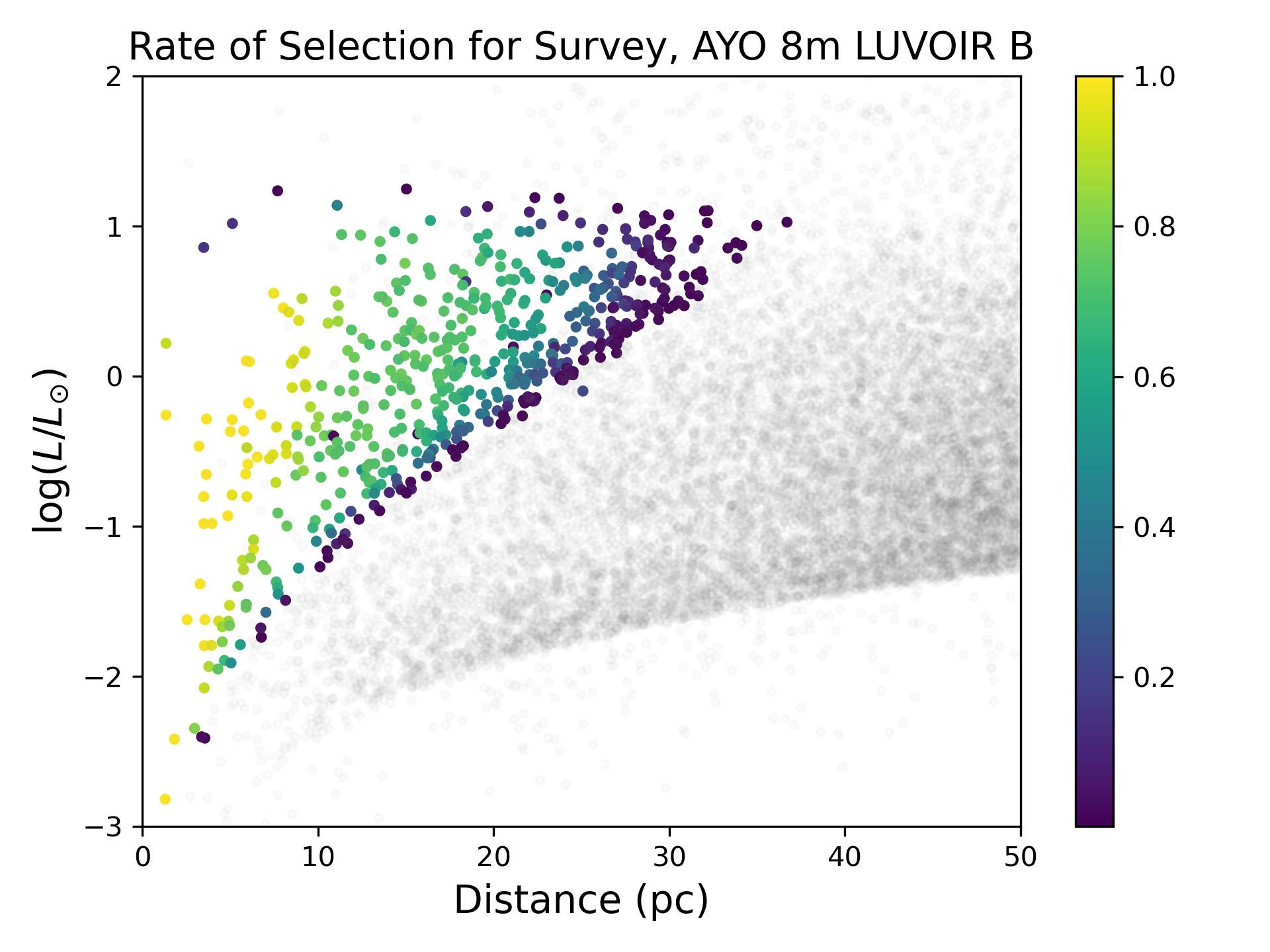}{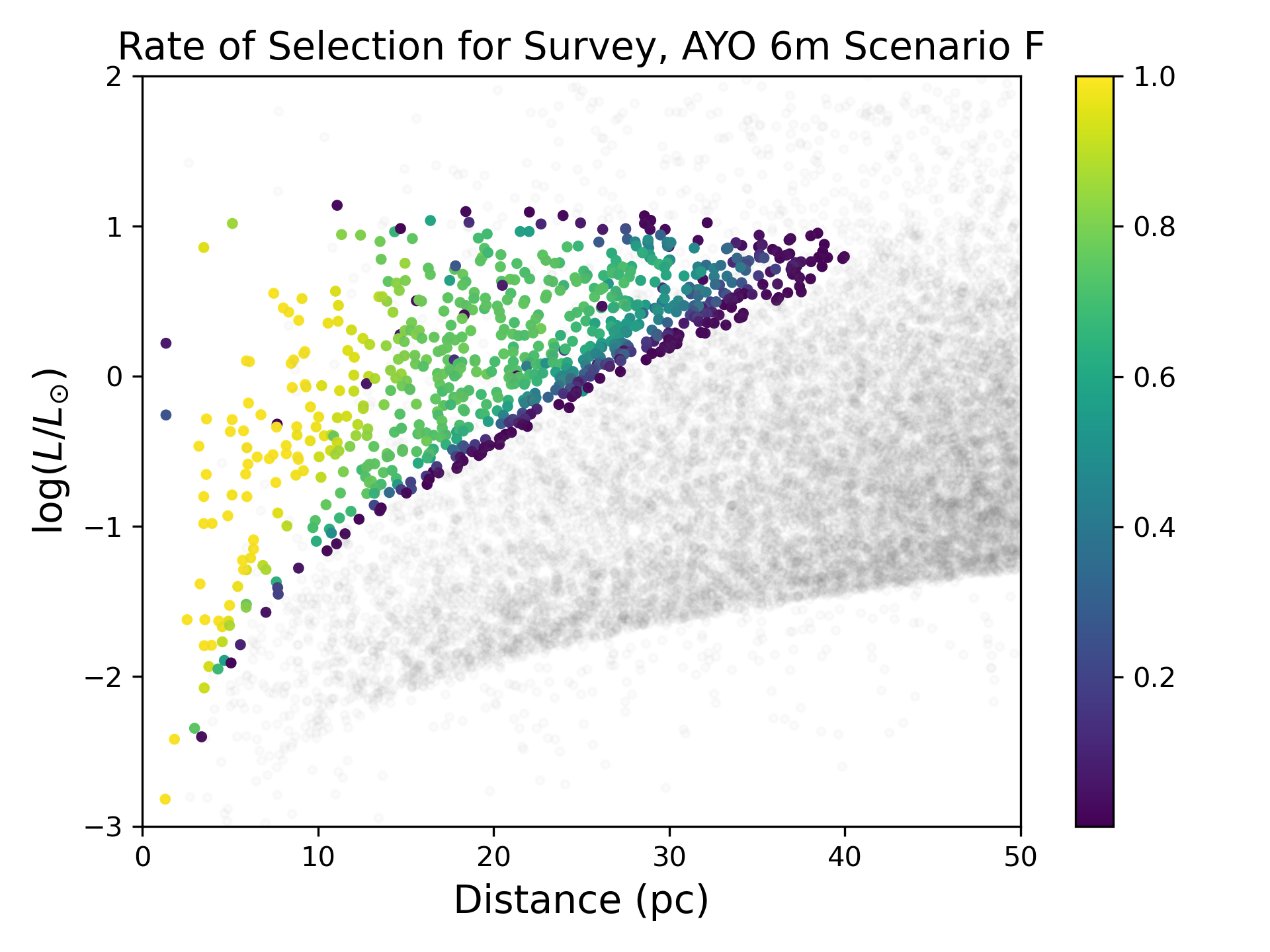}
    \caption{Stars selected by the AYO \citep{Stark24} 8m LUVOIR B yield calculation (left) and the 6m "scenario F" yield calculation (right) in distance vs. luminosity space. Points are color coded based on the fraction of simulations in which each star was selected. Gray points in the background are the stars in the HPIC \citep{Tuchow24}.}
    \label{AYO_yield_results_fig}
\end{figure*}

The first set of yield calculations our group used were the results of \citet{Stark24} using the AYO yield code. \citet{Stark24} focused on how astrophysical uncertainties in the inputs and assumptions used in yield calculations affect the mission output. 
They identified potential changes to the {HWO} mission design that could greatly improve exoEarth yield to ensure that a 6m mission would have a high probability of meeting the goal of characterizing 25 exoEarths.
We made use of data from two specific design scenarios from \citet{Stark24} encompassing edge cases for potential HWO designs. Using these edge cases allowed us to identify the overlap between their selected stars and constrain the likely population of HWO targets.
The first scenario was an 8m LUVOIR-B mission design. \citet{Stark24} used LUVOIR-B, a direct imaging mission concept submitted to the Astro2020 Decadal Survey which served as a direct inspiration for HWO, as its baseline design \citep{LUVOIR_final_report}. They investigated the effects of scaling the inscribed diameter of the telescope between 6 to 9 meters. We used their data for an 8 meter design, representing the largest diameter of HWO currently under consideration. 
The other scenario we used from this study was referred to as the 6m ``Scenario F" case.
This is a yield calculation exploring different methods to vastly improve the yield from a 6 meter telescope, including
for example: optimizing mirror coatings, model-based PSF subtraction, energy resolving detectors, high throughput coronagraphs, etc. Starting with the baseline LUVOIR B design, \citet{Stark24} progressively applied a series of design changes to the mission and observed how they influenced the exoEarth yield. Scenario F was the culmination of all these design trades, improving yield by a factor of 2.8 over the baseline design. 

Both of these design scenarios contained the results from 996 unique iterations of the yield simulation, accounting for stochasticity between runs based on random number generation. The output from these calculations provides the sample of stars selected for survey and how many iterations out of the 996 each star was selected for (i.e. the rate of selection). This stellar sample is illustrated in Figure \ref{AYO_yield_results_fig}, where stars are color coded by the fraction of simulations they were selected to be observed in. 
In distance and luminosity space this stellar population is bounded by the coronagraph inner working angle and required exposure time (right boundary) and an upper limit on stellar luminosity corresponding to the minimum achievable planet-star contrast (top boundary). One can see that the most frequently selected targets consist of the nearest and brightest stars as one would expect. The differences in mission design between the two scenarios affect the population of stars that they will observe, such as influencing the maximum distance of stars that will be selected. {One would typically expect a larger diameter telescope to be able to survey a larger sample of more distant stars, as coronagraph inner working angle scales as $\lambda/D$. However, for these design scenarios we see that the opposite is true, such that the design trades for the 6m scenario F sufficiently reduce the exposure time required per target to an extent that it offsets the advantages of having a larger mirror diameter.}

\begin{figure*}
    \centering
    \plotone{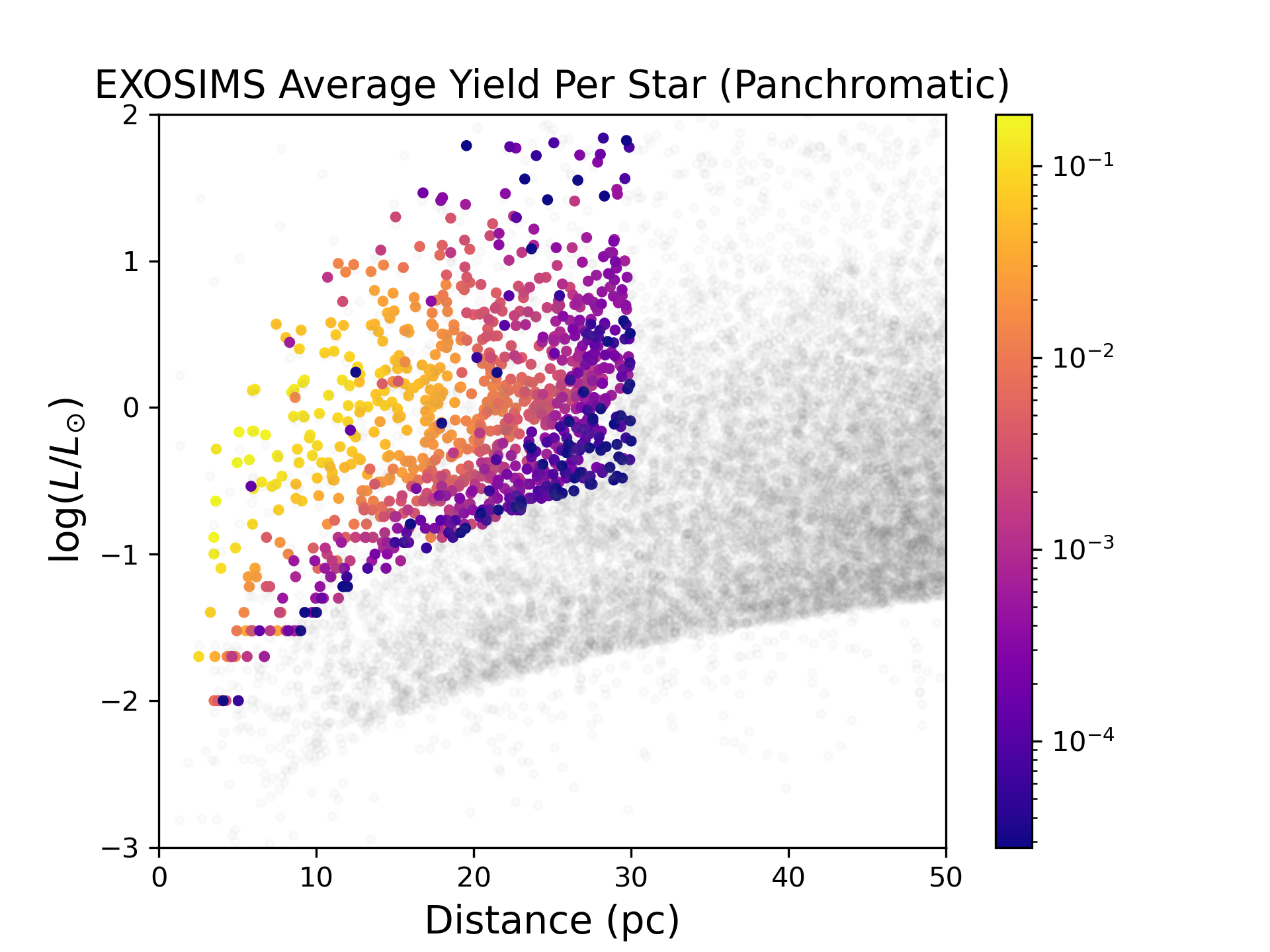}
    \caption{Stars selected by the ensemble of EXOSIMS calculations of \citet{Morgan24} in distance vs. luminosity space. Points are color coded based on the yield of exoEarths per star, averaged over an ensemble of scenarios, and averaged over the yield in the visible, IR, and UV bands to define a ``panchromatic yield metric". Gray points in the background are the stars in the HPIC. Note that these calculations used Exocat as an input catalog rather than the HPIC, which had a distance cutoff at 30 pc instead of the 50 pc for the HPIC.}
    \label{EXOSIMS_yield_results_fig}
\end{figure*}

The second major yield study that informed our construction of Tier 2 was the work of \citet{Morgan24} using the EXOSIMS yield code. This study focuses on the importance of detecting and characterizing exoEarths over the entire anticipated HWO wavelength range. 
Other yield studies have focused on specific subsets of the HWO bandpass. For instance, \citet{Stark24} considered visible wavelength detection of planets and characterization at 1 $\mathrm{\mu m}$.  \citet{Morgan24} instead focus on the ability of planets to be detected and spectrally characterized in the near infrared, optical, and near UV, and {identify} whether planets could be detected in all three bands. Unlike the yield calculations we used from \citet{Stark24}, which consisted of edge cases for HWO's design, this study did a sweep of potential mission design parameters and quantified how they affect exoEarth yield in these different bandpasses. { Their parameter sweep\footnote{Their upper bound for telescope diameter is slightly larger than the largest value explored by the EACs 1 - 3, though from the preliminary findings from these architectures, the next iteration of EACs may extend to slightly larger diameters.} considered telescope diameters between 6 to 9 meters, coronagraph throughputs between 0.1 and 0.6, inner working angles from 20 to 70 mas, and minimum contrasts from $10^{-9}$ to $10^{-11}$. }

\citet{Morgan24} provided the CDTG with the results from their ensemble of yield calculations spanning a large parameter space of telescope and coronagraph design parameters. These results contains the average yields per star over 354 scenarios where the mission design had sufficient telescope diameter and coronagraph inner working angle to detect planets in the habitable zone.
For each design scenario, the yield calculation was repeated for 100 iterations to account for stochasticity. The results of these yield calculations provide the average yields for each star in the visible, IR and UV bandpasses, and it defines a ``panchromatic yield metric" representing the per-star yield  averaged over all the bandpasses. In Figure \ref{EXOSIMS_yield_results_fig} we show the sample of stars observed over this ensemble of yield calculations, color coded by the average panchromatic yield metric per star. Note that targets beyond 30 pc were not included in this yield calculation as it used the ExoCat input catalog which is limited to stars within 30 pc \citep{Turnbull15}. Nonetheless, we can infer from this figure that stars beyond that distance cutoff would have close to zero yield per star, and are very unlikely to be selected.

\subsubsection{Combining Target Lists}
Using the results of the yield calculations for the 8m LUVOIR B and 6m scenario F designs from \citet{Stark24} and the large parameter sweep from \citet{Morgan24}, we sought to identify the overlap between their samples of target stars. 
Simply taking the union of the target stars from all three yield datasets was not sufficient as we know that many of the objects selected by these yield simulations may be relatively poor HWO targets, only observed for a few iterations of a very ambitious design scenario. Furthermore, a union of all these target lists would be too large, exceeding 1000 objects, and we want to reduce the size of {this list of likely target stars} to be small enough to study individually.
As such, it makes sense to trim these datasets to exclude outlier objects that are unlikely to be observed and have minimal contribution to the net yield of a mission. 
    
We aim to make consistent cuts on all the yield datasets, but this is made more complicated by the difference in output properties provided by the different studies. The \citet{Stark24} simulations give the number of times a star was observed over multiple iterations of the yield calculation, while the \citet{Morgan24}  calculations provide the number of scenarios that observed a target and its per star yield in different bandpasses. 
To identify the overlapping parameter space containing stars that were frequently selected for survey by all three yield calculations, the CDTG made the following cuts to eliminate outliers and reducing the size of the overall sample. For the EXOSIMS parameter sweep, we select the top 90\% of objects in terms of average panchromatic yield, sorting stars by yield and cutting the sample based on the cumulative distribution function. This was motivated by the shape of the distribution of yields per star, where there is a long tail of low yield targets that have minimal contribution to the net yield of the mission. For the 8m LUVOIR B and 6m scenario F datasets we make an analogous cut, selecting all objects that appear in more than 10\% of the 996 simulations. We then take the union of these three trimmed lists {and construct Tier 2 by selecting the stars that aren't already in Tier 1.} 
The final sample of Tier 2 targets {contains 495 stars and is shown in}  Figure \ref{all_tiers_plot}. It is important to note that the HWO mission is unlikely to observe all of these targets. Rather, a specific HWO design will observe some subset of this sample, and the purpose of including a large number of stars in Tier 2 is to identify the targets that are likely to be selected for survey, considering a wide range of possible HWO designs.

\subsection{Tier 3} 

Tier 3 of our target list contains all potential HWO targets {that are not included in the earlier tiers.}
{To construct Tier 3 we selected all objects in the HPIC catalog of \citet{Tuchow24} that are not in Tiers 1 or 2.}
{The HPIC makes minimal assumptions about telescope and coronagraph design choices so that it can be used as an input for yield calculations and other trade studies to assess the performance of proposed HWO designs. }
{It} was constructed by taking the union of the TESS Input Catalog and Gaia DR3 source catalog \citep{Stassun19,gaiadr3}, applying a distance cutoff at 50 pc. It applied a magnitude limit of 12 in Gaia $G$ and TESS $T$ bands to constrain the sample size without removing any feasible HWO targets. The resulting input catalog consists of 12,944 nearby, bright stars and, for each object, an automated pipeline gathers measurements and estimates of stellar properties from a wide variety of literature sources. This allows the list to be easily updated and reconstructed when new data become available.
This pipeline focuses primarily on stellar properties necessary for yield calculations such as distances, magnitudes, luminosities, radii, and stellar multiplicity.

The HPIC is used as an input for yield calculations simulating specific HWO mission architectures \citep[e.g.][]{Tuchow24,Stark24}. These yield calculations assess the quality of stars as direct imaging targets, and select a much smaller sample to be surveyed. Of the roughly 13,000 stars in the HPIC, it is likely that only $\sim$200 - 400 targets would be selected for survey when considering HWO architectures with inscribed diameters between 6 - 8 m (see Table 3 of \citealt{Tuchow24}). This implies that the vast majority of stars in Tier 3 are unlikely to be observed by a given HWO design{. For} these stars one should prioritize obtaining {coarse estimates of their stellar properties that are sufficient to achieve accurate yield estimates, rather than expending resources to obtain high precision measurements for stars that are unlikely to be selected by the exoEarth survey.}

\section{Contents of the HWO Target Stars and Systems 2025 List}

\label{data_products}
\begin{figure*}
    \centering
    \plotone{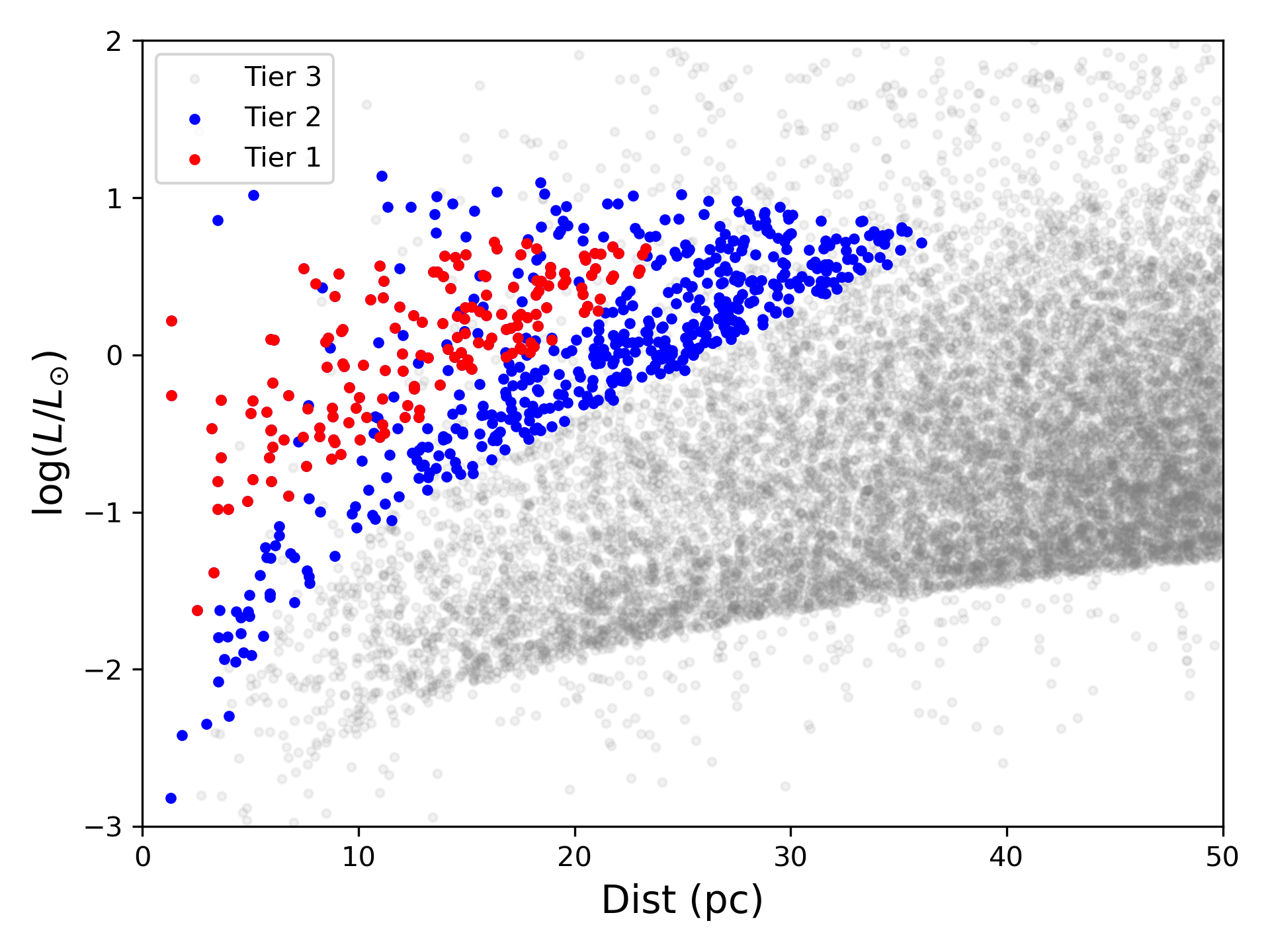}
    \caption{Targets in all tiers of our catalog plotted in distance vs. luminosity space, with the tiers indicated by color.}
    \label{all_tiers_plot}
\end{figure*}

\begin{figure*}
    \centering
    \plotone{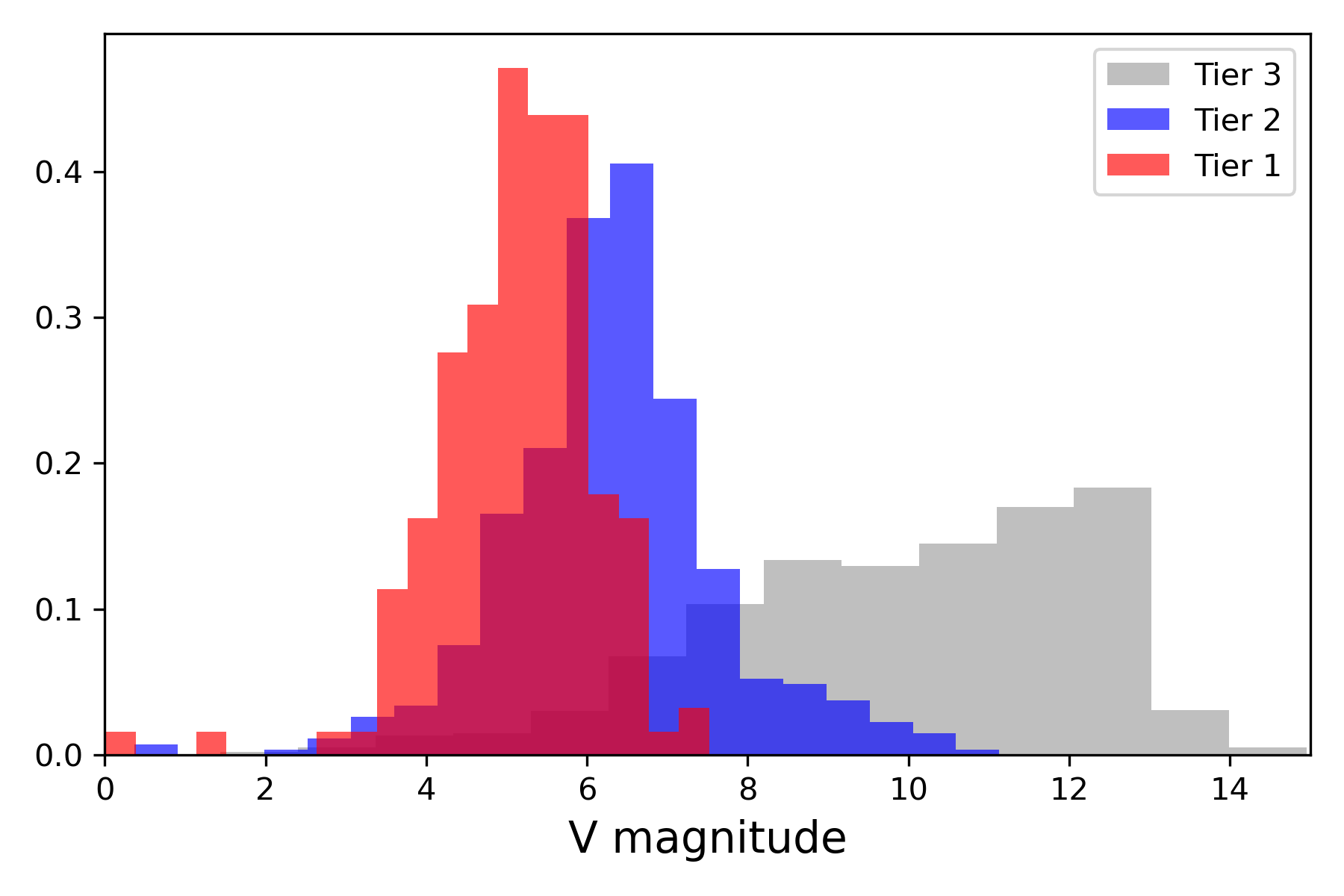}
    \caption{Histogram of Johnson V band magnitudes for stars in each tier of the TSS25 list.}
    \label{Vmag_hist}
\end{figure*}

The CDTG constructed the TSS25 list to identify which stellar targets are most likely to be observed by HWO, categorizing them into three priority tiers. 
A summary of the contents of each tier in the TSS25 catalog is provided in Figure \ref{all_tiers_plot}, with the number of stellar targets in each tier shown in Table \ref{tiers_table}. Figure \ref{all_tiers_plot} illustrates the distribution of stellar targets in distance and luminosity space, color coded according to their priority tier. Moving towards higher priority tiers, target stars occupy a narrower and narrower region of parameter space. In a histogram of V band magnitudes, we can see that objects in the higher tiers are on average brighter than those in the lower tiers (see Fig \ref{Vmag_hist}).
Tier 1 consists of the most accessible targets for the detection of exoEarths via direct imaging, and occupies a region of parameter space corresponding to the minimum required exposure time and the least demanding contrast and coronagraph inner working angle requirements. These consist primarily of the closest and brightest FGK-type dwarfs in the solar neighborhood with an average V magnitude of around 5.0. As we relax the inner working angle and contrast requirements, we get Tier 2 of our target list which extends to further distances and a wider range of stellar luminosities. This tier includes all objects with a high probability of being selected for observation by HWO regardless of HWO's telescope and coronagraph design. We constructed Tier 2 using the results of yield calculations that spanned the range of architectures for HWO currently under consideration. While a given HWO design will not survey all the stars in Tier 2, we can be confident that all the targets for such an HWO design will be included in {Tiers 1 and 2}. One may notice that the region of the plot corresponding to main-sequence FGK stars within 25 pc, which is occupied by the majority Tier 1 targets (in red), also contains a few Tier 2{ and 3} objects (in blue {and gray}). These objects likely host binary companions or other obstacles to direct imaging such that they weren't included in Tier 1, but {those in Tier 2} may be considered by more ambitious {mission} designs. 
Finally, Tier 3 consists of all nearby bright objects, subject to a distance and magnitude cutoff. It encompasses nearly the entire parameter space of this plot, subject only to the occurrence rates of objects in the solar neighborhood and a cutoff at 12th magnitude in the Gaia $G$ and TESS $T$ bands, forming the rough bottom curve for the gray points. In the histogram of V magnitudes (which are close but not equivalent to $G$ and $T$ mags), one observes the distribution of stars increases towards higher magnitudes until around 12.5, roughly corresponding to the limiting magnitude in the other bandpasses, indicating that this sample is volume-complete. This tier is more expansive than what is necessary for the stellar targets of currently considered HWO designs, but it is required as an input catalog for yield calculations which determine the performance of potential architectures and inform the higher tiers. 

An overview of the contents of the TSS25 target list can be found in Table \ref{columns_table}.
The most basic information contained within the TSS25 list is the primary object identifier (in this case the HPIC name) and the priority tier that it belongs to. It also contains cross-matching information for the identifiers of each object in several commonly used catalogs. In addition to these columns, it includes measurements of the positions, proper motions, parallaxes and V magnitudes of objects alongside  estimates of their distances and luminosities provided from the HPIC \citep{Tuchow24}. 

The TSS25 list of HWO targets is publicly available to the astronomical community at \url{https://zenodo.org/records/17195128}, and will be hosted on the CDS VizieR database. 
It was provided to the other task groups of the HWO Target Stars and Systems group, focusing on specific properties of host stars ranging from stellar multiplicity to UV fluxes to stellar activity indicators. These task groups worked on identifying what stellar observations are required for objects in the different priority tiers and quantifying the current state of our knowledge and the availability of measurements. They sought to identify the precursor science needs for HWO regarding the characterization of exoplanet host stars and determine which observations are required in advance of a mission. These groups plan to publish their findings in future reports, contributing additional data tables that can be cross-referenced with the TSS25 list, which serves as a master list between these different efforts.

The TSS25 list is a living catalog that will be updated as the state of our knowledge of these stars improves. 
The goal in the creation of the TSS25 list was to organize HWO precursor science efforts to identify which stars warrant the greatest focus of the community. With the construction of this target list and its priority tiers, greater attention can be directed toward objects in Tiers 1 and 2 which are likely to be observed by HWO, further refining estimates of their stellar properties and assessing their quality as direct imaging targets. While the sample of stars in Tiers 1 and 2 is larger than the sample that could be observed by a given HWO design, it is important to have an understanding of their stellar properties prior to the launch of a mission. To obtain a yield exceeding 25 exoEarths, a given mission design is likely to select most of the objects in Tier 1 plus several lower priority objects in Tier 2. These Tier 2 objects may require longer exposure times and individually have lower probabilities of hosting detectable exoEarths (on the order of 5-10\%), but observing many of these stars is required in order to meet the yield target of the mission. It is therefore critical that we don't only focus on the high priority objects in Tier 1, but also focus our observations on the objects in Tier 2 which also have a significant probability of being observed with HWO. Given that the design of HWO has yet to be finalized, Tier 2 is necessarily more expansive in order to take into account the range of possible architectures. Waiting until the design of HWO is finalized is not a viable option, as some precursor observations (such as radial velocity and astrometric searches for long period planets) will require upwards of a decade to complete, so they need to be started in the near future. {Currently there are multiple NASA funded proposals to better characterize the target stars for HWO, ranging from radial velocity searches for disruptive giant planets in the habitable zone, to constraining fundamental stellar properties using long-baseline interferometry, to characterizing the X-ray and UV emissions of target stars (see the 2022 and 2023 lists of selected proposals for the NASA ROSES Astrophysics Decadal Survey Precursor Science Program). However, several of these planned and ongoing observations only focus on a selection of the very best Tier 1 stars, and many of the likely HWO targets in Tiers 1 and 2 will be missing necessary precursor observations.}

The pros of observing and characterizing the sample of Tier 2 stars vastly outweigh the cons. If HWO launches and our precursor and preparatory science efforts have only focused on the Tier 1 stars, then there is a high probability that planets detected around Tier 2 stars will lack the necessary stellar observations, significantly delaying the interpretation of their spectra and the determination of their masses. Therefore precursor observations of Tier 2 stars are necessary to mitigate the risks of having insufficient information about these targets. While observing the entire sample of Tier 2 stars introduces the risk that several of these stars won't be observed by the mission, in the worst case we expend significant resources to learn more about about the properties of nearby stars in the solar neighborhood. HWO provides an exciting opportunity for astronomers to learn more about the nature of Earth-sized planets around Sun-like stars. Through identifying the target stars for HWO that should be the focus of precursor observations, we hope to ensure that HWO will have the necessary stellar information to understand the planets that will be discovered, maximizing its science return.

\section{Overview}

\begin{itemize}
    \item The HWO Target Stars and Systems 2025 (TSS25) list is community-curated list of potential targets for HWO's exoEarth direct imaging survey.
    
    \item It contains three priority tiers based on the likelihood of a star to be observed by HWO, {the} expected contribution to the mission's science output, and the importance of obtaining precursor observations.
    
    \item The TSS25 list is a living catalog that will be updated in future releases.
    
    \item This target list is publicly available\footnote{\url{https://zenodo.org/records/17195128}} and will be hosted on VizieR.
    
    \item We recommend that precursor studies focus their efforts on characterizing the 659 stars in Tiers 1 and 2 of the TSS25 list which have a high probability of being observed by proposed HWO designs. 
\end{itemize}

\begin{table*}[]
    \centering
    \caption{Columns in the TSS25 list}
    \begin{tabular}{l|l}
    \hline
    Column name & Description \\ \hline
    star\_name &  primary object identifier, given by HPIC name \\
    TSS\_tier  &  highest priority tier that an object belongs to \\
    ra &  right ascension at epoch J2000 (\degr)\\
    dec & declination at epoch J2000 (\degr) \\
    sy\_pmra &  proper motion in RA (mas/yr) \\
    sy\_pmdec & proper motion in DE (mas/yr) \\ 
    tic\_id &  TESS Input Catalog ID \\
    gaia\_dr3\_id &  Gaia DR3 ID \\
    hip\_name &  Hipparcos ID \\
    hd\_name &  Henry Draper Catalog ID \\
    tm\_name &  2MASS ID \\
    gj\_name & Gliese designation \\
    simbad\_name &  Name in the CDS Simbad database\\
    sy\_plx & parallax (mas) \\
    sy\_dist &  Distance (pc) \\
    st\_lum &  $\log_{10}$ stellar luminosity ($L_\odot$) \\
    sy\_vmag & Johnson V magnitude \\ \hline 
    \end{tabular}
    \label{columns_table}
\end{table*}

\section*{Acknowledgments}
\begin{acknowledgments}
This publication is a direct product of the HWO Target Stars and Systems sub-Working Group. The results and conclusions benefited greatly from group-wide discussions and analysis within the Catalogs \& Databases Task Group, as well as specific input from the Fundamental Properties Task Group, Activity \& Rotation Task Group, High Energy Emission Task Group, Multiple Stars Task Group, and EPRV Task Group.
Part of this research was carried out at the Jet Propulsion Laboratory, California Institute of Technology, under a contract with the National Aeronautics and Space Administration (80NM0018D0004).
NWT was supported by an appointment with the NASA Postdoctoral Program at the NASA Goddard Space Flight Center, administered by Oak Ridge Associated Universities under contract with NASA.
NWT and CCS acknowledge support from the Sellers Exoplanet Environments Collaboration (SEEC) at NASA's Goddard Space Flight Center. 
CKH acknowledges support from the National Science Foundation (NSF) Graduate Research Fellowship Program (GRFP) under Grant No.~DGE 2146752 and the NASA Astrophysics Decadal Survey Precursor Science (ADSPS) program under Grant No.~80NSSC23K1476.
The results reported herein benefited from collaborations and/or information exchange within NASA’s Nexus for Exoplanet System Science (NExSS) research coordination network sponsored by NASA’s Science Mission Directorate under Agreement No. 80NSSC21K0593 for the program ``Alien Earths".
\end{acknowledgments}

\bibliography{main}
\bibliographystyle{aasjournal}

\end{document}